\documentclass{acm_proc_article-sp}
\usepackage{caption}
\usepackage{subcaption}
\usepackage{verbatim}
\usepackage{listings}
\usepackage{url}

\newcommand{\figurename}{Figure}

\newcommand{\figref}[1]{\figurename~\ref{#1}}

\usepackage{tikz}
\usepackage{pgfplots}
\usetikzlibrary{arrows,chains,positioning,fit,backgrounds,calc,shapes,
  shadows,scopes,decorations.markings,plotmarks}

\newcommand*{\tettextsize}{\footnotesize}
\tikzstyle{line} = [draw, -, thick]
\tikzstyle{nodraw} = [draw, fill, circle, minimum width=0pt, inner sep=0pt]
\tikzstyle{sieve} = [line, circle, font=\tettextsize, inner sep=0pt,
  minimum size=12pt]

\tikzset{
  box/.style = {draw, thick, rectangle, font=\sf\scriptsize},
  file/.style = {
    box, fill=white, minimum size=2em, minimum height=3em
  },
  state/.style = {
    box, fill=black!20, minimum size=4em, minimum height=2em, rounded corners
  },
  multiple/.style = {
    double copy shadow={shadow xshift=1ex, shadow yshift=.6ex,
      draw=black!70}, draw=black, thick}
}

\newif\ifmonochrome
\monochromefalse

\ifmonochrome
\tikzstyle{cell} = [sieve]
\tikzstyle{facet} = [sieve]
\tikzstyle{edge} = [sieve]
\tikzstyle{vertex} = [sieve]
\else
\tikzstyle{cell} = [sieve, fill=blue!60]
\tikzstyle{facet} = [sieve, fill=green!35]
\tikzstyle{edge} = [sieve, fill=red!35]
\tikzstyle{vertex} = [sieve, fill=blue!35]
\fi

\graphicspath{{figures//}}

\title{Flexible, Scalable Mesh and Data Management using PETSc DMPlex}

\numberofauthors{3}
\author{
  \alignauthor
  Michael Lange\\
  \affaddr{Department of Earth Science and Engineering}\\
  \affaddr{Imperial College London, UK}\\
  \email{m.lange@imperial.ac.uk}
  \alignauthor
  Matthew G. Knepley\\
  \affaddr{Computation Institute}\\
  \affaddr{University of Chicago, USA}\\
  \email{knepley@gmail.com}
  \alignauthor
  Gerard J. Gorman\\
  \affaddr{Department of Earth Science and Engineering}\\
  \affaddr{Imperial College London, UK}\\
  \email{g.gorman@imperial.ac.uk}
}

\date{23 July 2015}

\begin{document}

\maketitle

\begin{abstract}
Designing a scientific software stack to meet the needs of the next-generation
of mesh-based simulation demands, not only scalable and efficient mesh and
data management on a wide range of platforms, but also an abstraction layer that makes it
useful for a wide range of application codes. Common utility tasks,
such as file I/O, mesh distribution, and work partitioning, should be delegated to
external libraries in order to promote code re-use, extensibility and
software interoperability. In this paper we demonstrate the use of
PETSc's DMPlex data management API to perform mesh input and domain
partitioning in Fluidity, a large scale CFD application. We demonstrate that
raising the level of abstraction adds new functionality to the
application code, such as support for additional mesh file formats and
mesh reordering, while improving simlutation startup cost through more
efficient mesh distribution. Moreover, the separation of concerns accomplished
through this interface shifts critical performance and interoperability issues, such
as scalable I/O and file format support, to a widely used and supported open source
community library, improving the sustainability, performance, and functionality
of Fluidity.

\end{abstract}

\keywords{Mesh, topology, partitioning, renumbering, Fluidity, PETSc}

\section{Introduction}

Scalable file I/O and efficient domain topology management present
important challenges for many scientific applications if they are to
fully utilise future exascale computing resources. Although these
operations are common to many scientific codes they have received
little attention in optimisation efforts, resulting in
potentially severe performance bottlenecks for realistic simulations
that require and generate large data sets. Moreover, due to a
multitude of formats and a lack of convergence on standards for mesh and output data
in the community there is only limited interoperability and very
little code reuse among scientific applications for common operations,
such as reading and partitioning input meshes. Thus developers are
often forced to create custom I/O routines or even use
application-specific file formats, which further limits application
portability and interoperability.

Designing a scientific software stack to meet the needs of the next-generation of
simulation software technologies demands, not only scalable and efficient
algorithms to perform data I/O and mesh management at scale, but also
an abstraction layer that allows a wide variety of application codes
to utilise them and thus promotes code reuse and
interoperability. Such an intermediate representation of mesh topology
has recently been added to PETSc~\cite{Balay2014}, a widely used scientific library for
the scalable solution of partial differential equations, in the form
of the DMPlex data management API~\cite{Knepley2009}.

In this paper we demonstrate the use of PETSc's DMPlex API to perform
mesh input and domain topology management in
Fluidity~\cite{Piggot2008}, a large scale CFD application code that
already uses the PETSc library as its linear solver engine. By
utilising DMPlex as the underlying mesh management abstraction we not
only add support for new mesh file formats, such as Exodus
II~\cite{schoof1994exodus}, CGNS~\cite{poirier1998cgns},
Gmsh~\cite{Geuz2009}, Fluent Case~\cite{FluentManual} and
MED~\cite{med}, to Fluidity, but also enable the use of domain
decomposition methods, data migration, and mesh renumbering techniques
at run-time. Moreover, the separation of concerns allows PETSc
parallel data management and HDF5 support to be independently
optimized for the target platform, removing this complexity from the
application code. Our refactoring of Fluidity provides significant
performance benefits due to improved cache locality and mesh
distribution during simulation initialisation, which we demonstrate
with performance benchmarks performed on Archer, a Cray XC30
architecture.

\section{Background}

The key challenge in designing software for large scale systems
lies in the composition of abstractions and the definition of clearly
defined yet flexible interfaces between them.
Code reuse and inter-disciplinary cooperation necessitate deeper
software stacks and thus configuration and extensibility will play a
key role in designing the software stack of the
future~\cite{Brown2015}. In this paper we therefore focus on the
interaction between applications and their supporting libraries to
provide the infrastructure for efficient data management at exascale.

\subsection{Fluidity}

The primary user application in our work is Fluidity, an open source
unstructured finite element code that uses mesh adaptivity to
accurately represent a wide range of scales in a single numerical
simulation without the need for nested grids. Fluidity is used in a
number of different scientific areas including geophysical fluid
dynamics, computational fluid dynamics, ocean modelling and mantle
convection. Fluidity implements various finite element and finite
volume discretisation methods and is capable of solving solving the
Navier-Stokes equation and accompanying field equations in one, two
and three dimensions.

Previous optimisation efforts have highlighted that file I/O, in
particular during model initialisation, presents a severe performance
bottleneck when running on large numbers of
processes~\cite{Guo2015}. The primary reasons for this are a off-line
domain partitioning and the need to store each partition using a
file-per-process strategy.

\subsection{DMPlex}

\begin{figure}[ht]
  \centering
  \begin{subfigure}{0.24\textwidth}\centering
    \begin{tikzpicture}[scale=1.2]
      \node (0) [nodraw, label=below:{\tettextsize 2}] at (0,0) {};
\node (1) [nodraw, label=below:{\tettextsize 3}] at (2.4,0) {};
\node (2) [nodraw, label=above right:{\tettextsize 4}] at (2.3,0.84) {};
\node (3) [nodraw, label=above:{\tettextsize 1}] at (1.2,2.0) {};

\path[line] (0) edge node[label=below:{\tettextsize 9}]{} (1);
\path[line, dashed] (0) edge node[label=above:{\tettextsize 14}]{} (2);
\path[line] (1) edge node[label=right:{\tettextsize 12}]{} (2);
\draw[line] (0) edge node[label=above left:{\tettextsize 11}]{} (3);
\draw[line] (1) edge node[label=left:{\tettextsize 10}]{} (3);
\draw[line] (2) edge node[label=above right:{\tettextsize 13}]{} (3);
    \end{tikzpicture}
    \label{fig:tet_simple}
    \caption*{Vertex and edge numbering}
  \end{subfigure}\begin{subfigure}{0.24\textwidth}\centering
    \begin{tikzpicture}[scale=1.2]
      \def\y{.79}
\def\x{.32}
\node (0) [cell] at (0,0) {0};
\node (1) [facet] at (-3*\x, \y) {5};
\node (2) [facet] at (-1*\x, \y) {6};
\node (3) [facet] at (1*\x, \y) {7};
\node (4) [facet] at (3*\x, \y) {8};
\node (5) [edge] at (-4*\x, 2*\y) {9};
\node (6) [edge] at (-2.4*\x, 2*\y) {10};
\node (7) [edge] at (-.8*\x, 2*\y) {11};
\node (8) [edge] at (.8*\x, 2*\y) {12};
\node (9) [edge] at (2.4*\x, 2*\y) {13};
\node (10) [edge] at (4*\x, 2*\y) {14};
\node (11) [vertex] at (-3*\x, 3*\y) {1};
\node (12) [vertex] at (-1*\x, 3*\y) {2};
\node (13) [vertex] at (1*\x, 3*\y) {3};
\node (14) [vertex] at (3*\x, 3*\y) {4};

\draw[line] (0) -- (1);
\draw[line] (0) -- (2);
\draw[line] (0) -- (3);
\draw[line] (0) -- (4);
\draw[line] (1) -- (5);
\draw[line] (1) -- (6);
\draw[line] (1) -- (7);
\draw[line] (2) -- (6);
\draw[line] (2) -- (8);
\draw[line] (2) -- (9);
\draw[line] (3) -- (7);
\draw[line] (3) -- (9);
\draw[line] (3) -- (10);
\draw[line] (4.north west) -- (5.south east);
\draw[line] (4) -- (8);
\draw[line] (4) -- (10);
\draw[line] (5) -- (12);
\draw[line] (5) -- (13);
\draw[line] (6) -- (11);
\draw[line] (6) -- (13);
\draw[line] (7) -- (11);
\draw[line] (7) -- (12);
\draw[line] (8) -- (13);
\draw[line] (8) -- (14);
\draw[line] (9) -- (11);
\draw[line] (9) -- (14);
\draw[line] (10) -- (12);
\draw[line] (10) -- (14);
    \end{tikzpicture}
    \caption*{Topological connectivity}
    \label{fig:tet_numbering}
  \end{subfigure}
  \caption{DAG-based representation of a single tetrahedron in
    DMPlex.}
  \label{fig:plex_overview}
\end{figure}
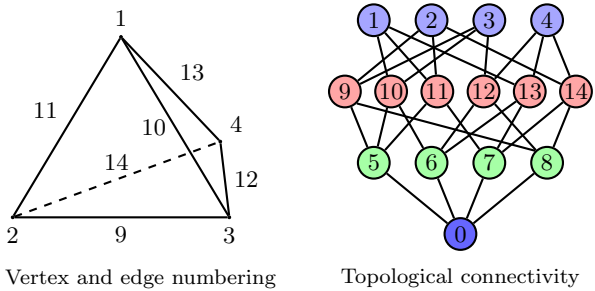

PETSc's ability to handle unstructured meshes is centred around
DMPlex, a data management object that encapsulates the topology of
unstructured grids to provide a range of functionalities common to
many scientific applications. As shown in \figref{fig:plex_overview},
DMPlex stores the connectivity of the associated mesh as a layered
directed acyclic graph (DAG), where each layer (stratum) represents a
class of topological entities~\cite{Knepley2009,Logg2009}. This
flexible yet efficient representation provides an abstract interface
for the implementation of mesh management and manipulation algorithms
using dimension-independent programming.

DMPlex stores data by associating data with points in the DAG, allowing
an arbitrary data size for each point. This can be efficiently encoded
using the same AIJ data structure used for sparse matrices. This scheme
is general enough to encompass any discrete data layout over a mesh. The
association with points also means that data can be moved using the same
set of scalable primitives that are used for mesh distribution.

DMPlex's internal representation of mesh topology also provides an
abstraction layer that decouples the mesh from the underlying file
format and thus allows support for multiple mesh file formats to be
added generically. At the time of writing DMPlex is capable of reading
input meshes in Exodus II, CGNS, Gmsh, Fluent-Case and MED
formats. Moreover, DMPlex provides output routines that generate
solution output in HDF5-based XDMF format, while also storing the
DMPlex DAG connectivity alongside the visualisable solution data to
facilitate checkpointing~\cite{Balay2014}.

In addition to a range of I/O capabilities DMPlex also provides
parallel data marshalling through automated parallel distribution of
the DMPlex~\cite{Knepley2015} and the pre-allocation of parallel
matrix and vector data structures. Mesh partitioning is provided via
internal interfaces to several partitioner libraries (Chaco,
Metis/ParMetis) and data migration is based on PETSc's internal {\em
  Star Forest} communication abstraction
(PetscSF)~\cite{Balay2014}. Additionally, DMPlex is designed to
provide the connectivity data and grid hierarchies required by
sophisticated preconditioners, such as geometric multigrid methods and
``Fieldsplit'' preconditioning for multi-physics problems, to speed up
the solution process~\cite{Brown2012,Brune2013}.

\subsection{Mesh Reordering}

Mesh reordering techniques represent a powerful performance
optimisation that can be utilised to increase cache coherency of the
matrices required during the solution
process~\cite{Gunther2006,Haase2007,Yoon2005}. The well-known
Reverse Cuthill-McKee (RCM) algorithm, which can be used to reduce the
bandwidth of CSR matrices, is implemented in PETSc allowing DMPlex to
compute the required permutation of mesh entities directly from the
domain topology DAG. The resulting permutation can then be applied to
any discretisation derived from the stored mesh topology to improve
the cache coherency of the associated CSR matrices.

\section{Fluidity-DMPlex Integration}

Initial mesh input has been a scalability bottleneck in Fluidity due to
the off-line mesh partitioning step. As illustrated in
\figref{fig:flplex-original}, the current preprocessor module
uses Zoltan~\cite{Zoltan06}, which use ParMetis~\cite{KarypisKumar97}
for graph partitioning, to partition and distribute the initial
simulation state to the desired number of processes before writing the
partitioned mesh and data to disk, allowing the main simulation to read the
pre-partitioned data in a parallel fashion.

Fluidity's parallel mesh initialisation routines, however, rely on a
file-per-process I/O strategy that require large numbers of individual
files when running the application at scale. This has been shown to
put significant pressure on the metadata servers in distributed
filesystems, such as Lustre or PVFS, which ultimately has a detrimental
effect on scalability when using sufficiently large numbers of processes~\cite{Guo2015}.

\subsection{Parallel Simulation Start-up}

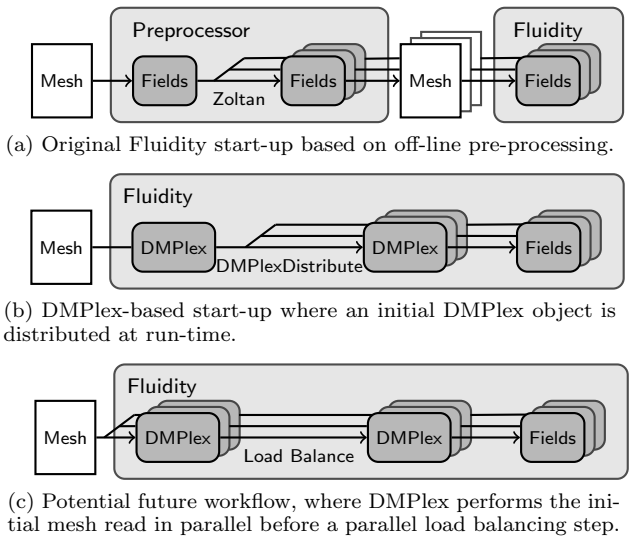
\begin{figure}[htp]\centering
  \vspace{1em}
  \begin{subfigure}{0.46\textwidth}
    \centering
    \begin{tikzpicture}[node distance=1.6em]
  \node(gmsh) [file] {Mesh};
  \node(flrd_seq) [state, right=1.6em of gmsh, minimum width=1em] {Fields};
  \node(flrd_par) [state, multiple, right=3.4em of flrd_seq.east, minimum width=1em] {Fields};
  \node(gmsh_par) [file, multiple, right=2.2em of flrd_par.east, minimum width=1em] {Mesh};
  \node(fluidity) [state, multiple, right=2.2em of gmsh_par.east, minimum width=1em] {Fields};

  \begin{scope}[overlay, remember picture, box, rounded corners, draw=black!70]
    \filldraw[fill opacity=0.1]
    ([xshift=-8pt, yshift=18pt] flrd_seq.north west) --
    ([xshift=16pt, yshift=18pt] flrd_par.north east) --
    ([xshift=16pt, yshift=-6pt] flrd_par.south east) --
    ([xshift=-8pt, yshift=-6pt] flrd_seq.south west) --
    cycle;
    \node [above=2pt of flrd_seq.north, xshift=10pt] {\small\sf Preprocessor};
  \end{scope}

  \begin{scope}[overlay, remember picture, box, rounded corners, draw=black!70]
    \filldraw[fill opacity=0.1]
    ([xshift=-8pt, yshift=18pt] fluidity.north west) --
    ([xshift=16pt, yshift=18pt] fluidity.north east) --
    ([xshift=16pt, yshift=-6pt] fluidity.south east) --
    ([xshift=-8pt, yshift=-6pt] fluidity.south west) --
    cycle;
    \node [above=2pt of fluidity.north] {\small\sf Fluidity};
  \end{scope}

  \draw[line, ->] (gmsh) -- (flrd_seq);
  \draw[line, ->] (flrd_seq) -- coordinate[pos=.2] (par) (flrd_par)
  node[midway, below=1pt] {\scriptsize\sf Zoltan};
  \draw[line] (par) -- ++(0.4,0.3) coordinate[pos=.5] (par1) coordinate[pos=1.] (par2);
  \draw[line] (par1) -- (flrd_par.161);
  \draw[line] (par2) -- (flrd_par.145) -- ++(0.08,0.);
  \draw[line, ->] (flrd_par) -- (gmsh_par);
  \draw[line] (flrd_par.19)+(.12,0) -- (gmsh_par.160);
  \draw[line] (flrd_par.34)+(.25,0) -- (gmsh_par.144);
  \draw[line, ->] (gmsh_par) -- (fluidity);
  \draw[line] (gmsh_par.20)+(.12,0) -- (fluidity.161);
  \draw[line] (gmsh_par.36)+(.25,0) -- (fluidity.145) -- ++(0.08,0);

\end{tikzpicture}
    \caption{Original Fluidity start-up based on off-line
      pre-processing.}
    \label{fig:flplex-original}
  \end{subfigure}
  \vspace{2em}

  \begin{subfigure}{0.46\textwidth}
    \centering
    \begin{tikzpicture}
  \node(gmsh) [file] {Mesh};
  \node(plex_seq) [state, right=1.6em of gmsh.east, minimum width=1em] {DMPlex};
  \node(plex_par) [state, multiple, right=6em of plex_seq.east, minimum width=1em] {DMPlex};
  \node(fluidity) [state, multiple, right=2.8em of plex_par.east, minimum width=1em] {Fields};

  \begin{scope}[overlay, remember picture, box, rounded corners, draw=black!70]
    \filldraw[fill opacity=0.1]
    ([xshift=-8pt, yshift=18pt] plex_seq.north west) --
    ([xshift=16pt, yshift=18pt] fluidity.north east) --
    ([xshift=16pt, yshift=-6pt] fluidity.south east) --
    ([xshift=-8pt, yshift=-6pt] plex_seq.south west) --
    cycle;
    \node [above=2pt of plex_seq.north, xshift=-6pt] {\small\sf Fluidity};
  \end{scope}

  \draw[line] (gmsh) -- (plex_seq);
  \draw[line, ->] (plex_seq) -- coordinate[pos=.2] (par) (plex_par)
  node[midway, below=1pt] {\scriptsize\sf DMPlexDistribute};
  \draw[line] (par) -- ++(0.4,0.3) coordinate[pos=.5] (par1) coordinate[pos=1.] (par2);
  \draw[line] (par1) -- (plex_par.165);
  \draw[line] (par2) -- (plex_par.152) -- ++(0.08,0.);
  \draw[line, ->] (plex_par) -- (fluidity);
  \draw[line] (plex_par.15)+(.12,0) -- (fluidity.161);
  \draw[line] (plex_par.28)+(.25,0) -- (fluidity.145) -- ++(0.08,0.);

\end{tikzpicture}
    \caption{DMPlex-based start-up where an initial DMPlex object is
      distributed at run-time.}
    \label{fig:flplex-current}
  \end{subfigure}
  \vspace{2em}

  \begin{subfigure}{0.46\textwidth}
    \centering
    \begin{tikzpicture}
  \node(gmsh) [file] {Mesh};
  \node(plex_seq) [state, multiple, right=1.6em of gmsh.east, minimum width=1em] {DMPlex};
  \node(plex_par) [state, multiple, right=6em of plex_seq.east, minimum width=1em] {DMPlex};
  \node(fluidity) [state, multiple, right=2.8em of plex_par.east, minimum width=1em] {Fields};

  \begin{scope}[overlay, remember picture, box, rounded corners, draw=black!70]
    \filldraw[fill opacity=0.1]
    ([xshift=-8pt, yshift=18pt] plex_seq.north west) --
    ([xshift=16pt, yshift=18pt] fluidity.north east) --
    ([xshift=16pt, yshift=-6pt] fluidity.south east) --
    ([xshift=-8pt, yshift=-6pt] plex_seq.south west) --
    cycle;
    \node [above=2pt of plex_seq.north, xshift=-6pt] {\small\sf Fluidity};
  \end{scope}

  \draw[line, ->] (gmsh) -- coordinate[pos=.2] (par) (plex_seq);
  \draw[line] (par) -- ++(0.4,0.3) coordinate[pos=.5] (par1) coordinate[pos=1.] (par2);
  \draw[line] (par1) -- (plex_seq.165);
  \draw[line] (par2) -- (plex_seq.152) -- ++(0.08,0.);
  \draw[line, ->] (plex_seq) -- (plex_par) node[midway, below=1pt,
    xshift=2pt] {\scriptsize\sf Load Balance};
  \draw[line] (plex_seq.15)+(.12,0) -- (plex_par.165);
  \draw[line] (plex_seq.28)+(.25,0) -- (plex_par.152) -- ++(0.08,0.);
  \draw[line, ->] (plex_par) -- (fluidity);
  \draw[line] (plex_par.15)+(.12,0) -- (fluidity.161);
  \draw[line] (plex_par.28)+(.25,0) -- (fluidity.145) -- ++(0.08,0.);

\end{tikzpicture}
    \caption{Potential future workflow, where DMPlex performs the
      initial mesh read in parallel before a parallel load balancing
      step.}
    \label{fig:flplex-target}
  \end{subfigure}

  \caption{Workflow diagram for Fluidity simulation start-up.}
  \label{fig:flplex}
\end{figure}

One of the objectives of this work, in addition to enhacing functionality
and usability, is to alleviate Fluidity's start-up
bottleneck by utilising DMPlex's mesh distribution capabilities to
perform mesh partitioning at run-time. For this purpose, as shown in
\figref{fig:flplex-current}, a DMPlex topology object is created from
the initial input mesh and immediately partitioned and distributed to
all participating processes, allowing Fluidity's initial coordinate
field to be derived from the DMPlex object in parallel. From the
initial coordinate mesh all further discretisations and fields in the
simulation state are then derived using existing functionality.

Using DMPlex as an intermediate representation for the underlying mesh
topology has the following advantages:
\begin{itemize}
\item Run-time mesh distribution and load balancing removes the need to store the
  partitioned mesh on disk, thus removing two costly I/O operations
  and reducing Fluidity's disk space requirements.
\item Communication volume during startup is reduced,
  since only the topology graph is distributed. This is in contrast to
  the preprocessor, which partitions and distributes a fully allocated
  Fluidity state with multiple fields.
\item Support for multiple previously unsupported mesh file formats is
  inherited from DMPlex, increasing application interoperability.
\end{itemize}

A key point to note about the DMPlex-based mesh initialisation
approach is that by delegating the initial mesh read to PETSc any mesh
format reader added to DMPlex in the future will automatically be
inherited by the application code. Moreover, future performance optimisations,
such as parallisation of the initial mesh file read, will also be
available in Fluidity without any further changes to the
application. Such an envisaged scenario is shown in
\figref{fig:flplex-target}, where an already parallel DMPlex object is
created from the initial file, followed by a load balancing step
before deriving the parallel Fluidity state.

\subsection{Mesh Renumbering}

One of the key components of the DMPlex integration is the derivation
of Fluidity's initial coordinate mesh object from the distributed
DMPlex, which includes the derivation of the data mapping required for
halo exchanges in parallel. DMPlex is able to provides such a mapping
from local non-owned degrees-of-freedom (DoFs) to their respective
remote owners. However, since Fluidity halo objects require remote
non-owned DoFs in the solution field to be located contiguously at the
end of the solution vector (``trailing receives'' assumption), a node
permutation is required when deriving Fluidity data structures from
the mapping provided by DMPlex.

\begin{figure}[htp]
  \centering
  \begin{subfigure}{0.23\textwidth}\centering
    \fbox{\includegraphics[width=0.9\textwidth, clip=true]
      {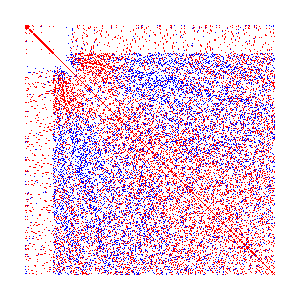}}
    \caption{Sequential, native}
  \end{subfigure}\begin{subfigure}{0.23\textwidth}\centering
    \fbox{\includegraphics[width=0.9\textwidth, clip=true]
      {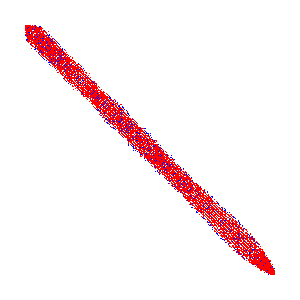}}
    \caption{Sequential, RCM}
  \end{subfigure}
  \vspace{4mm}

  \begin{subfigure}{0.23\textwidth}\centering
    \fbox{\includegraphics[width=0.9\textwidth, clip=true]
      {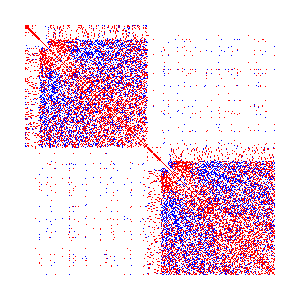}}
    \caption{2 MPI ranks, native}
  \end{subfigure}\begin{subfigure}{0.23\textwidth}\centering
    \fbox{\includegraphics[width=0.9\textwidth, clip=true]
      {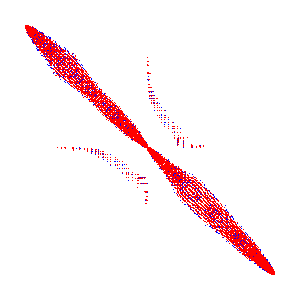}}
    \caption{2 MPI ranks, RCM}
  \end{subfigure}
  \caption{Effects of RCM reordering on the structure of the assembled
    pressure matrix.}
  \label{fig:matrix-reordering}
\end{figure}

As a consequence, further node ordering permutations may be applied
during the derivation of the initial field discretisation, such as the
RCM renumbering provided by DMPlex. An optional
renumbering step can be performed locally after the initial mesh
distribution and added to the mesh initialisation routine with very
little programming effort. As a result of Fluidity's run-time
derivation of field discretisations from the underlying coordinate
mesh, the RCM data layout of the initial reordering is inherited
by all fields in the simulation, as shown in
\figref{fig:matrix-reordering}. Moreover, as new mesh renumbering schemes
are incorporating into PETSc, they will be automatically available to the
application code.

\section{Results}

The following benchmark tests were performed on the UK national
supercomputer, a Cray XE30 with 4920 nodes connected via an
Aries interconnect and a parallel Lustre filesystem~\footnote{\url{http://www.archer.ac.uk/}}. Each node
consists of two 2.7 GHz, 12-core Intel E5-2697 v2 (Ivy Bridge)
processors with 64GB of memory.

\begin{figure}[ht]
  \centering
  \begin{subfigure}{0.23\textwidth}\centering
    \includegraphics[width=0.9\textwidth, clip=true]
      {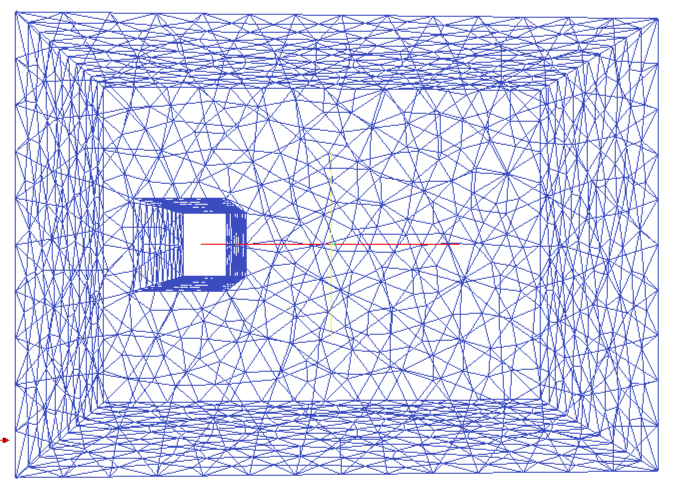}
  \end{subfigure}\begin{subfigure}{0.23\textwidth}\centering
    \includegraphics[width=0.9\textwidth, clip=true]
      {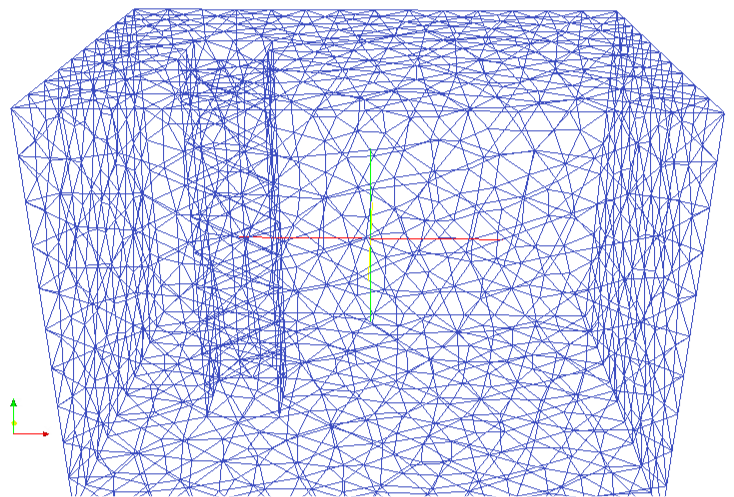}
  \end{subfigure}

  \caption{Three-dimensional benchmark mesh used to model flow past a cylinder.}
  \label{fig:mesh-3d-cylinder}
\end{figure}

The benchmark runs simulate the flow past a square cylinder for 10
timesteps using a $P^{\mathrm{DG}}_1-P_2$
discretisation~\cite{cotter2009mixed}, where a second-order pressure
field is solved using Fluidity's multigrid algorithm and paired with a
discontinuous first order velocity field that uses a GMRES solver with
SOR preconditioning. The mesh used has been generated with the Gmsh
mesh generator~\cite{Geuz2009} and is shown in
\figref{fig:mesh-3d-cylinder}.

\begin{figure}[htp]\centering
  \begin{subfigure}{0.5\textwidth}\centering
    \includegraphics
        [width=\textwidth, clip=true]
        {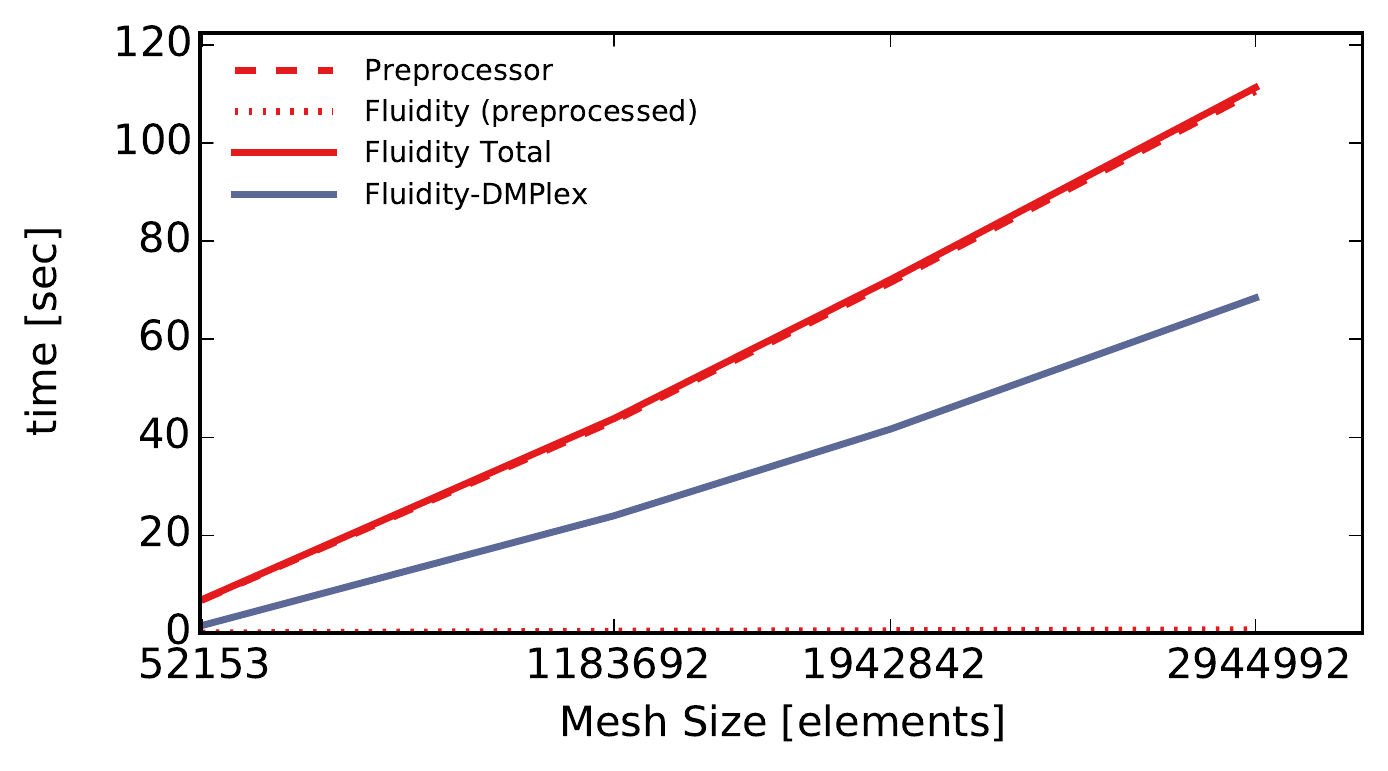}
        \caption{Overall simulation start-up}
        \label{fig:result-startup-total}
  \end{subfigure}

  \begin{subfigure}{0.5\textwidth}\centering
    \includegraphics
        [width=\textwidth, clip=true]
        {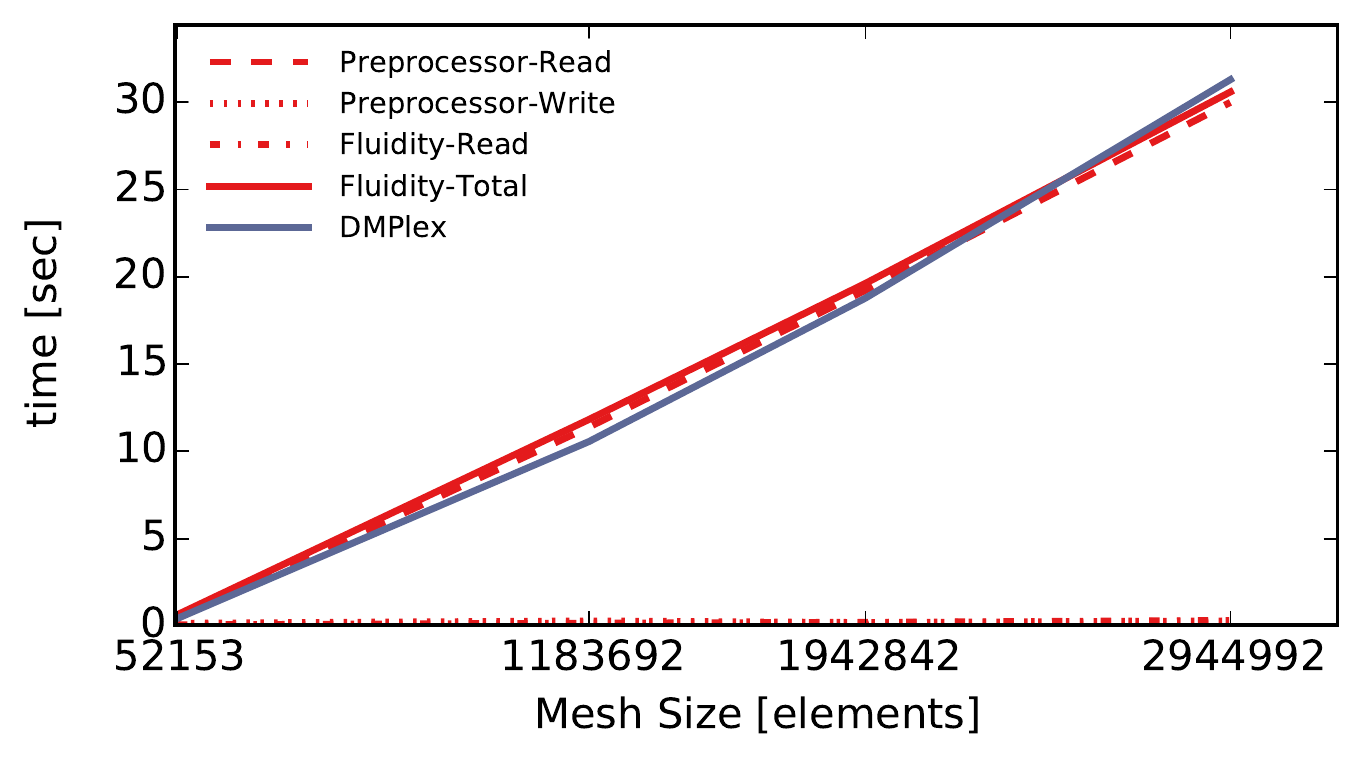}
        \caption{Mesh file I/O}
        \label{fig:result-startup-io}
  \end{subfigure}

  \begin{subfigure}{0.5\textwidth}\centering
    \includegraphics
        [width=\textwidth, clip=true]
        {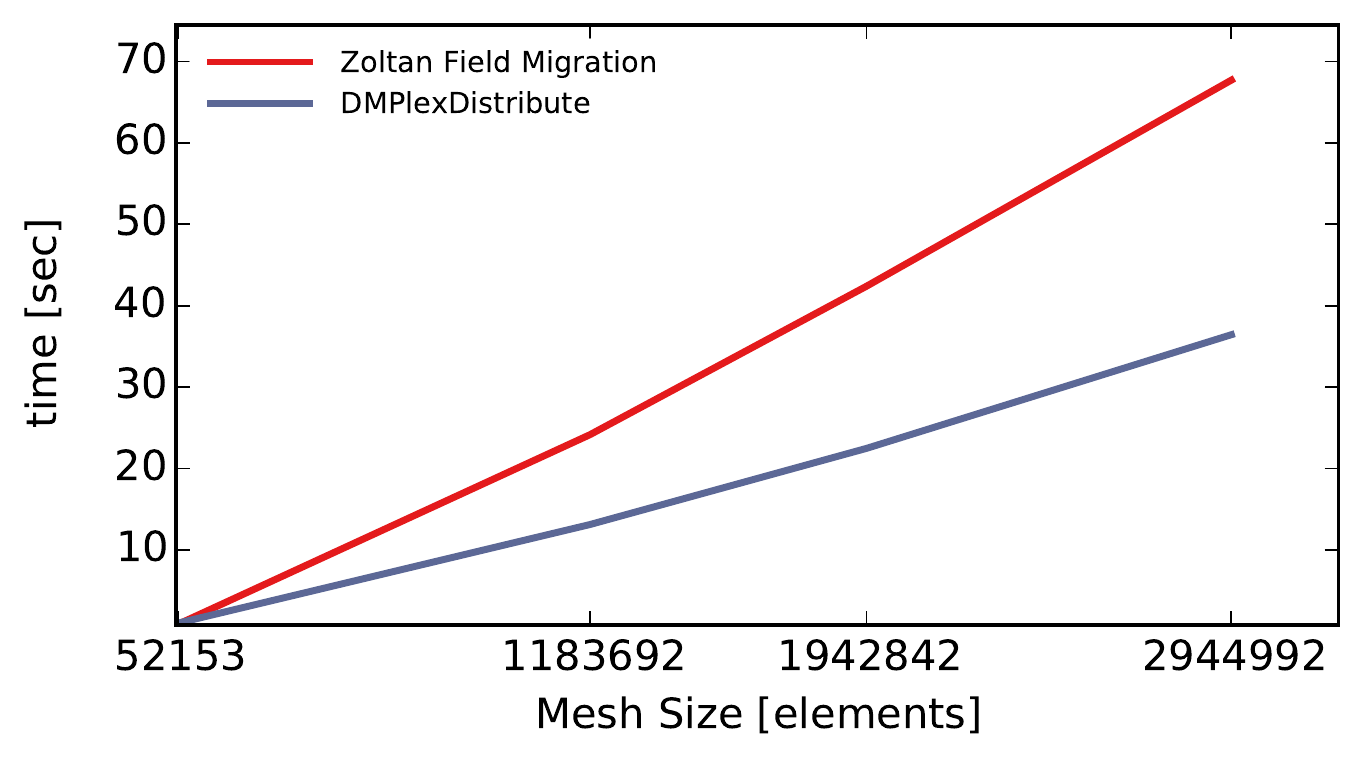}
        \caption{Mesh distribution}
        \label{fig:result-startup-distribute}
  \end{subfigure}

  \caption{Comparison of total start-up cost between preprocessor and
    DMPlex-based approaches.}
  \label{fig:result-startup}
\end{figure}

\begin{figure}[htp]\centering
  \begin{subfigure}{0.5\textwidth}\centering
    \includegraphics
        [width=\textwidth, clip=true]
        {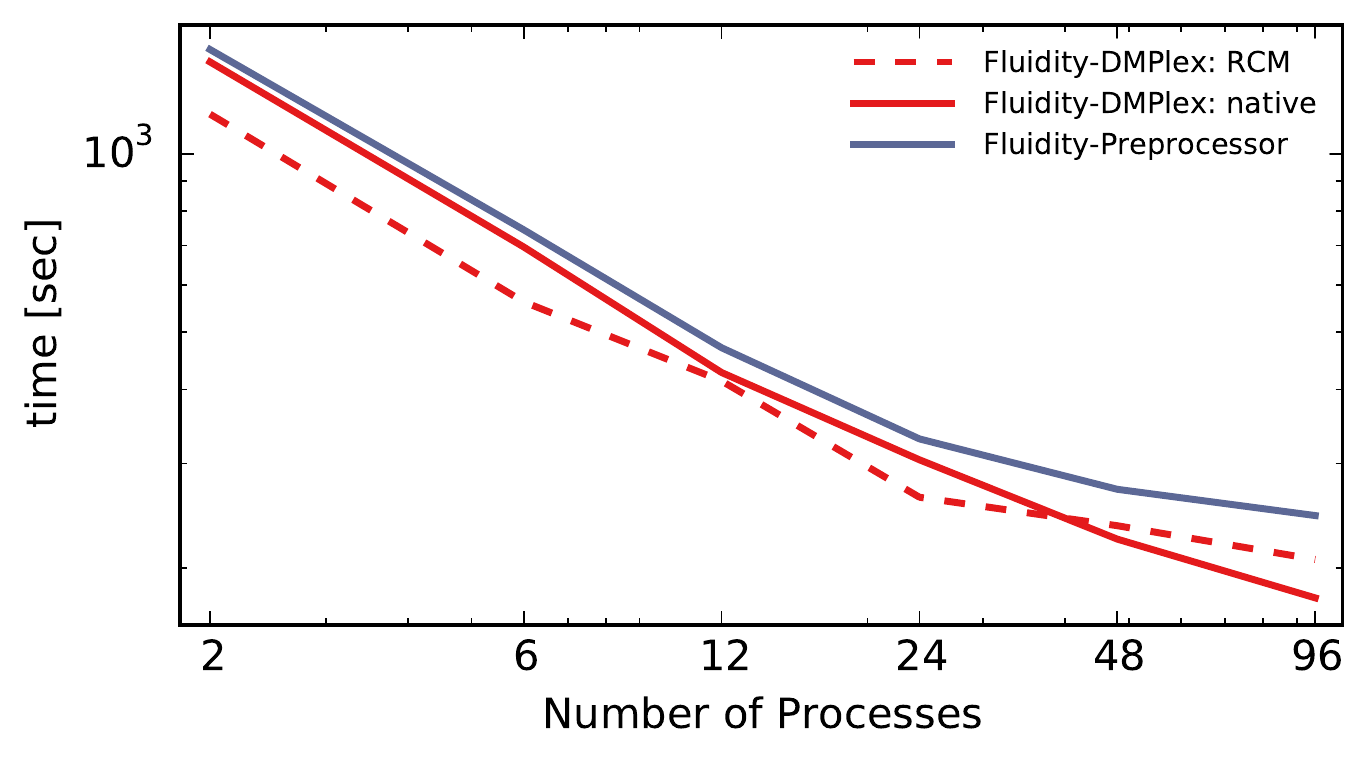}
        \caption{Overall simulation performance.}
        \label{fig:result-sim-total}
  \end{subfigure}

  \begin{subfigure}{0.5\textwidth}\centering
    \includegraphics
        [width=\textwidth, clip=true]
        {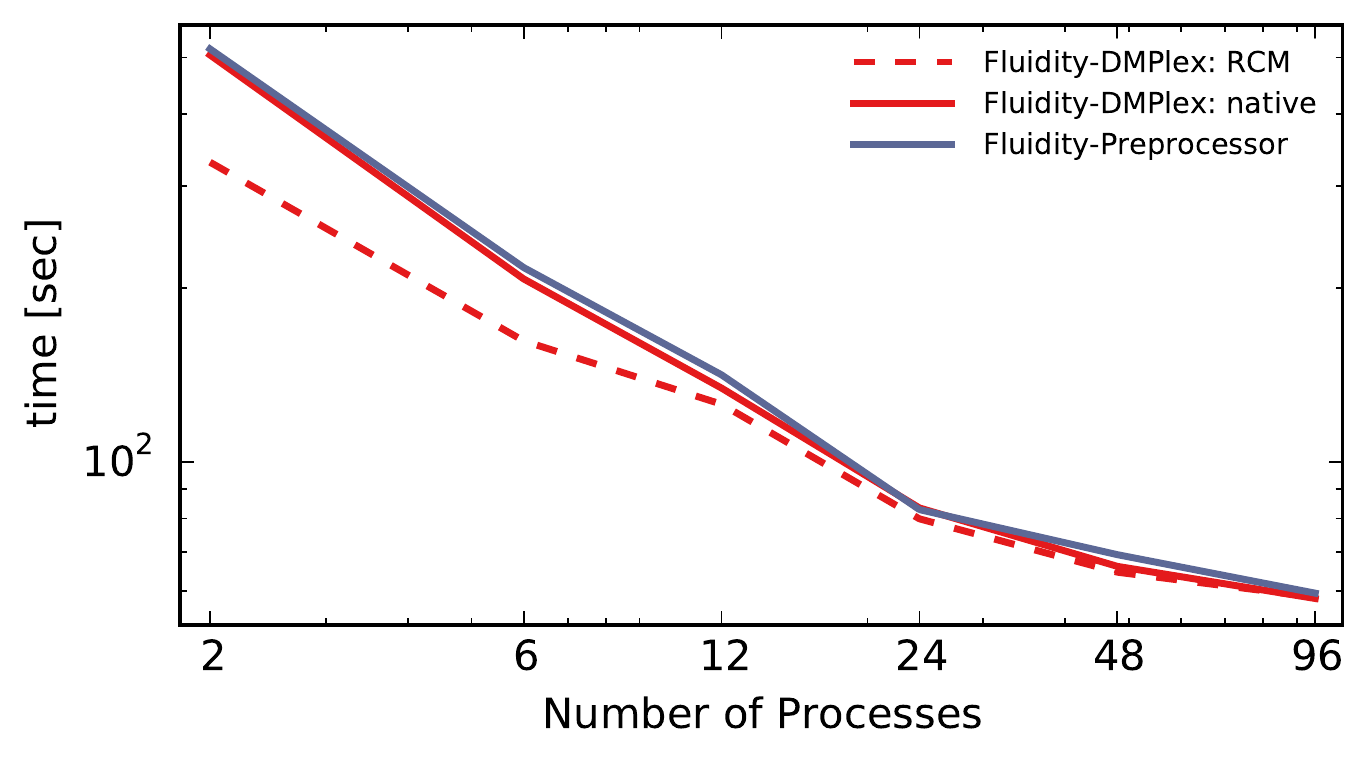}
        \caption{Pressure solve performance.}
        \label{fig:result-sim-pressure}
  \end{subfigure}

  \begin{subfigure}{0.5\textwidth}\centering
    \includegraphics
        [width=\textwidth, clip=true]
        {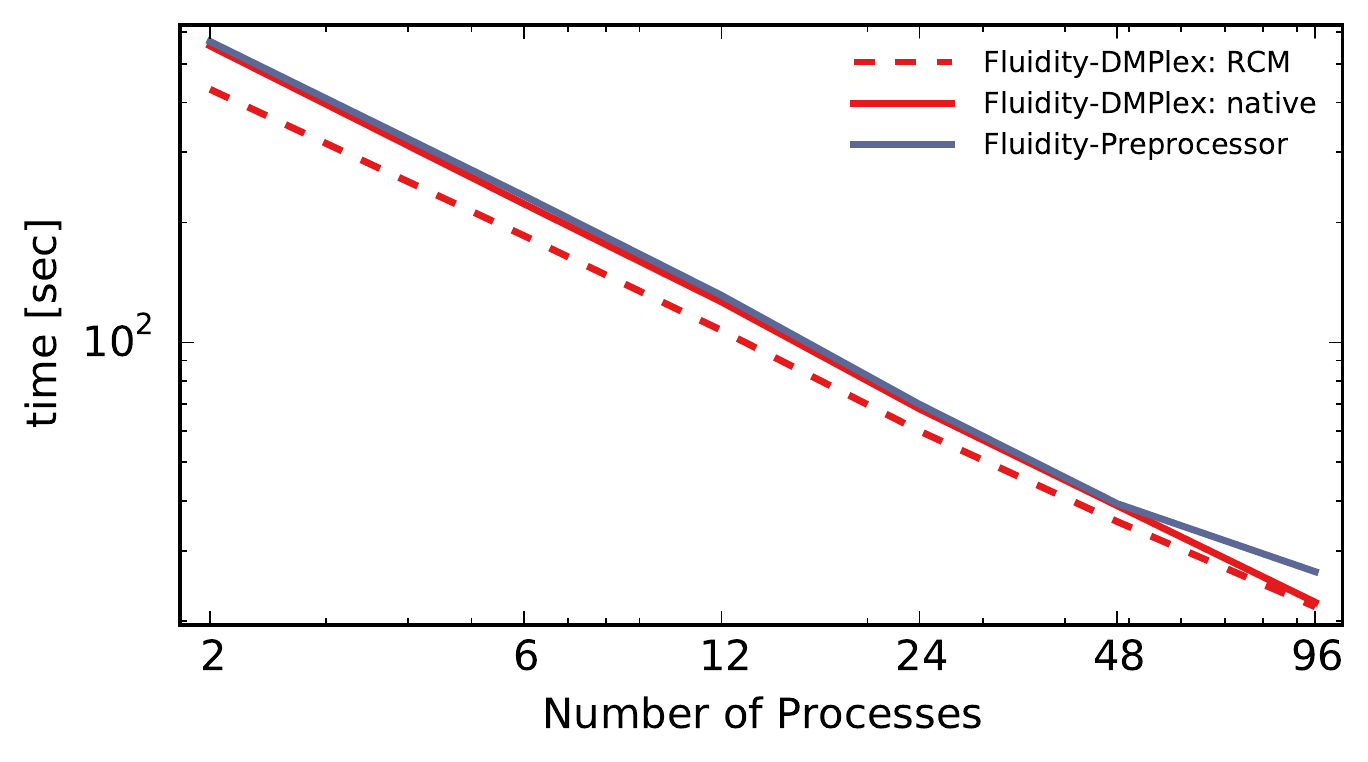}
        \caption{Velocity assembly performance.}
        \label{fig:result-sim-velocity}
  \end{subfigure}
  \caption{Full simulation performance for 10 timesteps on
    approximately 3 million elements.}
  \label{fig:result-sim}
\end{figure}

\subsection{Mesh Initialisation}

\figref{fig:result-startup} shows a comparison of the simulation
start-up cost between the DMPlex-based implementation and the original
preprocessor approach on 4 nodes (96 cores) with increasing mesh sizes
up to approximately 3 million elements (weak scaling). The original
start-up cost is hereby quantified as the sum of preprocessor and
simulation initialisation times.

A clear improvement in overall start-up performance is shown in
\figref{fig:result-startup-total}, although no significant improvement
in direct file I/O, as shown in \figref{fig:result-startup-io}, can be
determined. This is unsurprising, as file I/O is still completely
dominated by the initial sequential read, although potential gains can
be expected at larger scales due to removing the intermediate reads
and writes of the partitioned mesh.

As highlighted in \figref{fig:result-startup-distribute}, the majority
of the observed overall performance gains can be attributed to
significantly improved mesh distribution via DMPlex. It is important
to note here that DMPlex partitions and migrates only the mesh
topology graph and its associated coordinate values, in contrast to
the original preprocessor module that distributes fully assembled
fields. As a result less data needs to be communicated during the mesh
migration phase, resulting in significantly increased start-up
performance.

\subsection{Mesh Renumbering}

The overall simulation performance, including the effects of the mesh
reordering derived from DMPlex, are evaluated in
\figref{fig:result-sim}. This benchmark compares the performance of
both implementations by running 10 timesteps of the full simulation
using a mesh with approximately 3 million elements on up to 96
cores. The results, shown in \figref{fig:result-sim-total}, indicate a
consistent performance improvement of the DMPlex-based model with
native mesh ordering over the preprocessor approach that increases
with growing numbers of processes due to a smaller start-up overhead.

The effect of RCM mesh reordering is best demonstrated by analysing
the two most expensive components of the simulation: the pressure
field solve (\figref{fig:result-sim-pressure}) and the assembly of the
velocity matrix (\figref{fig:result-sim-velocity}). Both components
exhibit significant performance increases with RCM reordering on small
numbers of processors that diminish as the simulation approaches the
strong scaling limit. However, the benefits for overall simulation
performance (\figref{fig:result-sim-velocity}) with RCM reordering
decrease between 24 and 96 processes due to the fixed-cost start-up
overhead of generating the permutation outweighing the solver and
assembly benefits.

\section{Discussion}

Achieving scalable performance with production-scale scientific
applications on future exascale systems requires appropriate levels of
abstraction across the entire software stack. In this paper we report
progress on the integration of DMPlex, a library-level domain topology
abstraction, with the application code Fluidity in order to delegate a
set of common mesh and data management tasks to a widely used
library. We highlight the increased interoperability achieved through
the inheritance of new mesh file format readers and demonstrate
improved model initialisation performance through run-time mesh
distribution routines provided by DMPlex.

The key benefit of the restructured model initialisation workflow,
however, lies in the fact that responsibility for supporting various
mesh file formats and optimising mesh file I/O now lies with the
underlying library. This entails that any future extensions, such as
new file formats or parallel mesh reader implementations, are
automatically inherited by Fluidity, as well as other applications
using PETSc, such as Firedrake~\cite{Rathgeber2015} where we have also
employed these abstractions. Moreover, by utilising a centralised mesh management API
other types of mesh-based performance optimisations become available
to the application, as highlighted by the seamless addition of the RCM
renumbering feature.

\subsection{Future Work}

A key contribution of this work lies in the fact that it enables
future extensions and optimisations. Most crucially perhaps is the
development of a fully parallel mesh input reader in PETSc in order to
overcome the remaining sequential bottleneck during model
initialisation. This change, however, requires a new default mesh
format for Fluidity due to the inherently sequential nature of the
Gmsh file format, which again highlights the need for abstraction when
optimising mesh management.

In addition to the optimisation of mesh input and model initialisation,
further integration of DMPlex throughout Fluidity is desirable to
utilise DMPlex's advanced I/O features, such as the HDF5-based Xdmf
output format. For this purpose closer integration is required, where
additional discretisation data needs to be passed to PETSc to perform
all the necessary field I/O. Moreover, DMPlex's mesh and data
distribution utility may also be used to provide load balancing after
mesh adaptation.

\section{Acknowledgments}
This work was supported by the embedded CSE programme of the ARCHER UK
National Supercomputing Service \\(http://www.archer.ac.uk), and the Intel Parallel Computing
Center program through grants to both the University of Chicago and Imperial College London. We would also like to thank Frank Milthaler for
providing the test configurations used for benchmarking.

\bibliographystyle{abbrv}
\bibliography{references}

\end{document}